\documentclass[structabstract]{aa}  
                                   
\usepackage[hyperindex=true,colorlinks=true,citecolor=blue,linkcolor=blue,breaklinks=true]{hyperref}
\usepackage{amsmath}
\usepackage{graphicx}
\usepackage{lscape}
\usepackage{txfonts}
\usepackage{natbib,twoopt}
\usepackage{color}

\bibpunct{(}{)}{;}{a}{}{,} 
\newcommandtwoopt{\citeads}[3][][]{\href{http://adsabs.harvard.edu/abs/#3}%
{\citealp[#1][#2]{#3}}} 
\newcommandtwoopt{\citepads}[3][][]{\href{http://adsabs.harvard.edu/abs/#3}%
{\citep[#1][#2]{#3}}} 
\newcommandtwoopt{\citetads}[3][][]{\href{http://adsabs.harvard.edu/abs/#3}%
{\citet[#1][#2]{#3}}}
\newcommandtwoopt{\citeyearads}[3][][]%
{\href{http://adsabs.harvard.edu/abs/#3}{\citeyear[#1][#2]{#3}}}

\begin{document}

	\title{Searching for visual companions of close Cepheids}
	\subtitle{VLT/NACO lucky imaging of Y~Oph, FF~Aql, X~Sgr, W~Sgr and $\eta$~Aql\thanks{Based on observations made with ESO telescopes at Paranal observatory under program ID 089.D-0040}}
	
	\titlerunning{Searching for visual companions of close Cepheids}


	\author{ A.~Gallenne\inst{1} \and
				P.~Kervella\inst{2} \and
				A.~M\'erand\inst{3} \and
				N.~R.~Evans\inst{4} \and
				J.~H.~V.~Girard\inst{3} \and 
				W.~Gieren\inst{1}\and
				G.~Pietrzy\'nski\inst{1,5}	
  				}
  				
  	\authorrunning{A. Gallenne et al.}

\institute{Universidad de Concepci\'on, Departamento de Astronom\'ia, Casilla 160-C, Concepci\'on, Chile
	\and LESIA, Observatoire de Paris, CNRS UMR 8109, UPMC, Universit\'e
  Paris Diderot, 5 Place Jules Janssen, F-92195 Meudon, France
  	\and European Southern Observatory, Alonso de C\'ordova 3107,
  Casilla 19001, Santiago 19, Chile
  \and  Smithsonian Astrophysical Observatory, 
  MS 4, 60 Garden Street, Cambridge, MA 02138, USA
  \and Warsaw University Observatory, Al. Ujazdowskie 4, 00-478, Warsaw, Poland
  }
  
  \offprints{A. Gallenne} \mail{agallenne@astro-udec.cl}

   \date{Received March 24, 2014; accepted May 13, 2014}

 
  \abstract
  {}
   {High-resolution imaging in several photometric bands can provide color and astrometric information of the wide-orbit component of Cepheid stars. Such measurements are needed to understand the age and evolution of pulsating stars. In addition, binary Cepheids have the potential to provide direct and model-independent distances and masses.}
   {We used the NAOS-CONICA adaptive optics instrument (NACO) in the near-infrared to perform a deep search for wide components around the classical Cepheids, Y~Oph, FF~Aql, X~Sgr, W~Sgr, and $\eta$~Aql, within a field of view (FoV) of $1.7\arcsec \times 1.7\arcsec$ ($3.4\arcsec \times 3.4\arcsec$ for $\eta$~Aql).}
   {We were able to reach contrast $\Delta H = 5$--8\,mag and $\Delta K_\mathrm{s} = 4$--7\,mag in the radius range $r > 0.2\arcsec$, which enabled us to constrain the presence of wide companions. For Y~Oph, FF~Aql, X~Sgr, W~Sgr, and $\eta$~Aql at $r > 0.2\arcsec$, we ruled out the presence of companions with a spectral type that is earlier than a B7V, A9V, A9V, A1V, and G5V star, respectively. For $0.1\arcsec < r < 0.2\arcsec$, no companions earlier than O9V, B3V, B4V, B2V, and B2V star, respectively, are detected. A component is detected close to $\eta$~Aql at projected separation $\rho = 654.7 \pm 0.9$\,mas and a position angle $PA = 92.8 \pm 0.1\degr$. We estimated its dereddened apparent magnitude to be $m_H^0 = 9.34 \pm 0.04$ and derived a spectral type that ranges between an F1V and F6V star. Additional photometric and astrometric measurements are necessary to better constrain this star and check its physical association to the $\eta$~Aql system.}
   {}

 
 \maketitle

%

\section{Introduction}

Classical Cepheids are of fundamental importance for the Galactic and extragalactic distance scale. Therefore, a good understanding of their physical properties is necessary. More particularly, their mass and distance need to be known with a good accuracy to better constrain models of pulsating stars. Cepheids in binary systems are unique tools to independently estimate the dynamical mass and distance.

However, companions to Cepheids are hard to detect. Cepheids are bright-supergiant stars, outshining their less evolved companion, which is still on the main sequence. In addition, the small angular separation of most of the companions with respect to the Cepheid ($\sim 1-50$\,mas) makes their spatial resolution difficult with single-dish telescopes. So far, only four (wide) companions have been imaged with the Hubble Space Telescope \citep[Polaris, $\eta$~Aql, V659~Cen and S~Nor,][]{Evans_2008_09_0,Remage-Evans_2013_10_0} ; otherwise, the presence of the companions are generally revealed from UV spectra \citep[e.g.][]{Evans_1992_01_0} or radial velocity (RV) measurements \citep[e.g.][]{Szabados_1991_01_0}. Recent examples of the effectiveness of the radial velocity method to detect Cepheids in binary systems are the discoveries of three Cepheids in eclipsing binary systems in the LMC \citep{Pietrzynski_2010_11_0, Pietrzynski_2011_12_0, Pilecki_2013_12_0, Gieren_2014_05_0}. Imaging and RVs are complementary techniques. Indeed, only close companions have a notable effect on the Cepheid RVs, while wide components can be detected from imaging. The second difficulty lies in the brightness difference between the two stars. Most of the companions are hot main-sequence stars ; therefore the brightness of Cepheids makes their detection difficult for wavelengths longer than $0.5\,\mathrm{\mu m}$.

Recently, \citet{Gallenne_2013_04_0,Gallenne_2013_10_0} used multi-telescope interferometry to spatially resolve the companions located at a few milli-arcseconds (mas) from the Cepheids. Astrometric measurements combined with spectroscopy were used to determine all the orbital elements of the V1334~Cyg system. However, RV measurements of the secondary are still missing to enable mass and distance estimates. Although interferometry is complementary to the previous mentioned techniques that probe spatial scales $< 50$\,mas \citep[see also][]{Gallenne_2013_02_0}, it is still limited to bright companions (typically $\Delta H \lesssim 5-6$).

We report new observations of five Classical Cepheids using adaptive optic (AO) imaging in two broad- and two narrow-band filters. These observations allowed us to search for companions inside $1.7\arcsec \times 1.7\arcsec$  and $3.4\arcsec \times 3.4\arcsec$ field of views (FoV). The observations and data reduction procedure are detailed in Sect.~\ref{section__observation_and_data_reduction}. We then describe the method we used to search for companions in Sects.~\ref{section__psf_subtraction} and \ref{section__detection_limits}. Astrometric and photometric measurements of the companion of $\eta$~Aql is detailed in Sect.~\ref{section__the_case_of_eta_aql}. Our results are then discussed in Sect.~\ref{section__discussion}.

\onltab{
\begin{table*}
\centering
\caption{Log of our NACO observations.}
\begin{tabular}{ccccclccccccc}
\hline
\hline
\#  & JD	&	Star	& $\phi$ 	&	Filter\tablefootmark{a} 	&	FoV	 &	DIT	&	N	&	Seeing$_\lambda$	&	AM  & $r_0$	&	$t_0$	\\
 	 &			&				&				&						& 			& (ms)	&	&	(\arcsec)						&			&	(cm)	&	(ms)		\\
\hline
 1   & 2456126.5747	&	\object{Y~Oph}						& 0.31 &	NB1.64	& $1.7\arcsec\times1.7\arcsec$&	16	&	3000	&	0.65	&	1.09	&	19.1	&	6.4	\\
 2   & 2456126.5749	&	Y~Oph						& 0.31 &	$H$		& $1.7\arcsec\times1.7\arcsec$&	16	&	3000	&	0.63	&	1.09	&	19.6	&	6.4	\\
 3   & 2456126.5766	&	Y~Oph						& 0.31 &	NB2.17	& $1.7\arcsec\times1.7\arcsec$&	16	&	3000	&	0.58	&	1.09	&	20.3	&	6.5	\\
 4   & 2456126.5775	&	Y~Oph						& 0.31 &	$Ks$	& $1.7\arcsec\times1.7\arcsec$&	16	&	3000	&	0.59	&	1.09	&	20.0	&	6.1	\\
 5   & 2456126.5827	&	\object{HD~159527}				&	--	&	NB1.64	& $1.7\arcsec\times1.7\arcsec$&	16	&	3000	&	0.59	&	1.03	&	22.2	&	8.9		\\
 6   & 2456126.5836	&	HD~159527				&	--	&	$H$			&$1.7\arcsec\times1.7\arcsec$ &	16	&	3000	&	0.60	& 1.02		&	23.1	&	10.1	\\
 7   & 2456126.5846	&	HD~159527				&	--	&	NB2.17	  & $1.7\arcsec\times1.7\arcsec$&	16	& 3000	&	0.56	&	1.02	& 22.4	&	10.6	\\
 8   & 2456126.5855	&	HD~159527				&	--	&	$Ks$		& $1.7\arcsec\times1.7\arcsec$&	16	&	3000	&	0.57	&	1.02	&	22.6	&	11.0	\\
 9   & 2456127.7393 &	\object{FF~Aql}						& 0.66 &	NB1.64		& $1.7\arcsec\times1.7\arcsec$&	16	&	3000	&	0.88	&	1.57	&	12.9	&	4.4	\\
10  & 2456127.7402 &	FF~Aql						& 0.66 &	$H$			&$1.7\arcsec\times1.7\arcsec$ &	16	&	3000	&	0.83	&	1.57	&	12.7	&	4.2	\\
11  & 2456127.7411 &	FF~Aql						& 0.66 &	NB2.17		& $1.7\arcsec\times1.7\arcsec$&	16	&	3000	&	0.77	&	1.58	&	12.8	&	4.0	\\
 12 & 2456127.7420 &	FF~Aql						& 0.66 &	$Ks$		& $1.7\arcsec\times1.7\arcsec$&	16	&	3000	&	0.78	&	1.58	&	12.6	&	3.8	\\
 13 & 2456127.7473 &	\object{HD~175743}			&	--	&	NB1.64		& $1.7\arcsec\times1.7\arcsec$&	16	&	3000	&	0.60	&	1.65	&	10.9	&	4.1	\\
 14 & 2456127.7482 &	HD~175743				&	--	&	$H$			& $1.7\arcsec\times1.7\arcsec$&	16	&	3000	&	0.60	&	1.66	&	10.9	&	4.1	\\
 15 & 2456127.7491 &	HD~175743				&	--	&	NB2.17		& $1.7\arcsec\times1.7\arcsec$&	16	&	3000	&	0.56	&	1.67	&	11.2	&	4.4	\\
 16 & 2456127.7500 &	HD~175743				&	--	&	$Ks$		&$1.7\arcsec\times1.7\arcsec$ &	16	&	3000	&	0.54	&	1.68	&	11.3	&	4.4	\\
 17 & 2456128.7986 &	\object{$\eta$~Aql}					& 0.97 &	NB1.64	& $3.4\arcsec\times3.4\arcsec$&	39	&	2000	&	1.05	&	1.40	&	13.6	&	4.8	\\
 18 & 2456128.7999 &	$\eta$~Aql					& 0.97 & $H$			& $3.4\arcsec\times3.4\arcsec$&	39	&	2000	&	1.05	&	1.41	&	13.2	&	4.7	\\
19  & 2456128.8012 &	$\eta$~Aql					& 0.97 &	NB2.17	& $3.4\arcsec\times3.4\arcsec$&	39	&	2000	&	0.99	&	1.42	&	12.5	&	4.4	\\
 20 & 2456128.8025 &	$\eta$~Aql					& 0.97 &	$Ks$		& $3.4\arcsec\times3.4\arcsec$&	39	&	2000	&	0.98	&	1.43	&	12.4	&	4.4	\\
 21 & 2456128.8067 &	\object{HD~188512}				&	--	&	NB1.64	& $3.4\arcsec\times3.4\arcsec$&	39	&	2000	&	0.68	&	1.54	&	13.2	&	4.8	\\
 22 & 2456128.8079 &	HD~188512				&	--	&	$H$		& $3.4\arcsec\times3.4\arcsec$&	39	&	2000	&	0.72	&	1.55	&	13.1	&	4.8	\\
 23 & 2456128.8092 &	HD~188512				&	--	&	NB2.17	& $3.4\arcsec\times3.4\arcsec$&	39	&	2000	&	0.68	&	1.56	&	12.8	&	4.7	\\
24  & 2456128.8104 &	HD~188512				&	--	&	$Ks$	& $3.4\arcsec\times3.4\arcsec$&	39	&	2000	&	0.69	&	1.57	&	12.8	&	4.7	\\
 25 & 2456167.6426 &	\object{X~Sgr}						& 0.68 &	NB1.64			& $1.7\arcsec\times1.7\arcsec$&	16	&	3000	&	0.81	&	1.42	&	17.2	&	8.8	\\
26  & 2456167.6434 &	X~Sgr						& 0.68 &	$H$				& $1.7\arcsec\times1.7\arcsec$&	16	&	3000	&	0.71	&	1.43	&	17.4	&	8.9	\\
27  & 2456167.6444 &	X~Sgr						& 0.68 &	NB2.17		& $1.7\arcsec\times1.7\arcsec$&	16	&	3000	&	0.78	&	1.43	&	17.5	&	9.3	\\
28  & 2456167.6453 &	X~Sgr						& 0.68 &	$Ks$		& $1.7\arcsec\times1.7\arcsec$&	16	&	3000	&	0.73	&	1.44	&	17.5	&	9.5	\\
29  & 2456167.6694 &	\object{W~Sgr}						& 0.38 &	NB1.64			& $1.7\arcsec\times1.7\arcsec$&	16	&	3000	&	0.67	&	1.54	&	13.9	&	4.9	\\
30  & 2456167.6703 &	W~Sgr						& 0.38 &	$H$			& $1.7\arcsec\times1.7\arcsec$&	16	&	3000	&	0.67	&	1.55	&	14.5	&	5.2	\\
31  & 2456167.6712 &	W~Sgr						& 0.38 &	NB2.17		& $1.7\arcsec\times1.7\arcsec$&	16	&	3000	&	0.66	&	1.56	&	14.6	&	5.2	\\
32  & 2456167.6721 &	W~Sgr						& 0.38 &	$Ks$		& $1.7\arcsec\times1.7\arcsec$&	16	&	3000	&	0.67	&	1.57	&	14.4	&	4.9	\\
33  & 2456167.6817 &	\object{HD~166295}				&	--	&	NB1.64	& $1.7\arcsec\times1.7\arcsec$&	16	&	3000	&	0.87	&	1.65	&	12.7	&	5.2	\\
34  & 2456167.6827 &	HD~166295				&	--	&	$H$			&$1.7\arcsec\times1.7\arcsec$ &	16	&	3000	&	0.83	&	1.67	&	13.0	&	5.3	\\
35  & 2456167.6836 &	HD~166295				&	--	&	NB2.17		&$1.7\arcsec\times1.7\arcsec$ &	16	&	3000	&	0.71	&	1.68	&	13.0	&	5.6	\\
36  & 2456167.6845 &	HD~166295				&	--	&	$Ks$		& $1.7\arcsec\times1.7\arcsec$&	16	&	3000	&	0.77	&	1.69	&	13.2	&	5.6	\\
 \hline 
\end{tabular}
\tablefoot{$\phi$: pulsation phase, estimated with the ephemeris from \citet{Samus_2009_01_0}. N: number of frames in the cube. FoV: field-of-view. Seeing$_\lambda$: seeing at the observed wavelength, converted from the one measured in $V$ at the spots of the Shack-Hartmann active optics wavefront sensor. AM: mean airmass. $r_0, t_0$: average values of the Fried parameter and coherence time, estimated from the adaptive-optics real time calculator.\\
\tablefoottext{a}{All broad-band observations were used with a neutral density filter.} 
}
\label{table__log}
\end{table*}
}

\section{Observation and data reduction}
\label{section__observation_and_data_reduction}

The Cepheids were observed with the NACO instrument installed at the Nasmyth B focus of UT4 of ESO VLT. This is an adaptive optics system \citep[NAOS]{Rousset_2003_02_0} and a high-resolution near-IR camera \citep[CONICA]{Lenzen_2003_03_0}, which works as an imager or as a spectrograph in the range 1--$5\,\mathrm{\mu m}$. We used the S13 camera (FoV of $13.5\arcsec \times 13.5\arcsec$) with the $H$ and $Ks$ broadband filters ($\lambda_0 = 1.66\,\mathrm{\mu m}, \Delta \lambda = 0.33\,\mathrm{\mu m}$ and $\lambda_0 = 2.18\,\mathrm{\mu m}, \Delta \lambda = 0.35\,\mathrm{\mu m}$, respectively), and the narrow band filters NB\_1.64 ($\lambda_0 = 1.644\,\mathrm{\mu m}, \Delta \lambda = 0.018\,\mathrm{\mu m}$) and NB\_2.17 ($\lambda_0 = 2.166\,\mathrm{\mu m}, \Delta \lambda = 0.023\,\mathrm{\mu m}$). We chose the cube mode which allows us to record very short exposures, and reach truly diffraction-limited images. The efficiency of the cube mode vs. the standard long exposure mode is discussed in \citet{Kervella_2009_09_0}. A gain of 9\,\% in Strehl ratio with respect to a long exposure has been demonstrated by \citet{Girard_2010_07_0} in the $K$ band. For all observations in broadband, the neutral density filter \texttt{ND\_Short} (reducing the intensity by a factor of 80) was used due to the target brightness.

Data were obtained in July and August 2012 with a window size of $128 \times 130$ pixels, and the shortest allowed exposure time of 16\,ms, except for $\eta$~Aql for which a window size of $256 \times 258$ pixels and 39\,ms exposure time were used. The log of our observations is presented in Table~\ref{table__log}. The point spread function (PSF) calibrator stars were observed immediately after the Cepheids with the same instrumental configurations, except for X Sgr, for which the chosen calibrator had moved out too quickly from the FoV, for an unknown technical reason. For this star, we used the calibrator of W~Sgr observed one hour later. 

Each raw image in the cube was processed in a standard way, using bias subtraction, flat-field, and bad pixel corrections. The negligible sky background was not subtracted. We then carried out a precentering and a sorting according to the maximum intensity of the central peak. We then applied the shift-and-add technique \citep{Bates_1980_03_0} to obtain the best possible angular resolution. This method enables to enhance the Strehl ratio by selecting the frames that are the least altered by the atmospheric turbulence. When used with adaptive optics, this technique also reduces the halo contribution (i.e the residual light out of the coherent core) by selecting the best AO-corrected frames. Our processing steps were as follows. We first selected the top 10\,\% of the frames with the brightest pixel (as a tracer of the Strehl ratio) in our data cubes. We then spatially resampled them by a factor of 4 using a cubic spline interpolation and co-aligned them using a Gaussian fitting on the central core at a precision level of a few mas \citep[this method is described in detail in][]{Kervella_2009_09_0}. Each cube was then averaged to obtain the final mean images. 


\section{PSF subtraction}
\label{section__psf_subtraction}

For each final Cepheid image, we subtracted the corresponding PSF-reference image to remove most of the contribution from the central part. The classical PSF-subtraction technique significantly enhanced the capability to detect companions. We first recentered the PSF image, according to the Cepheid image using a cubic spline interpolation, and then scaled the PSF flux with respect to the Cepheid for $0 < r < 0.1\arcsec$ ($\sim 2\lambda / D$). We therefore expect to detect the companions farther than 0.1\arcsec. We finally cleaned the images by removing the correlated noise (correlated lines in the background detector) by performing a median subtraction per line and then a ring median subtraction.

\begin{figure*}
\centering
\resizebox{\hsize}{!}{\includegraphics{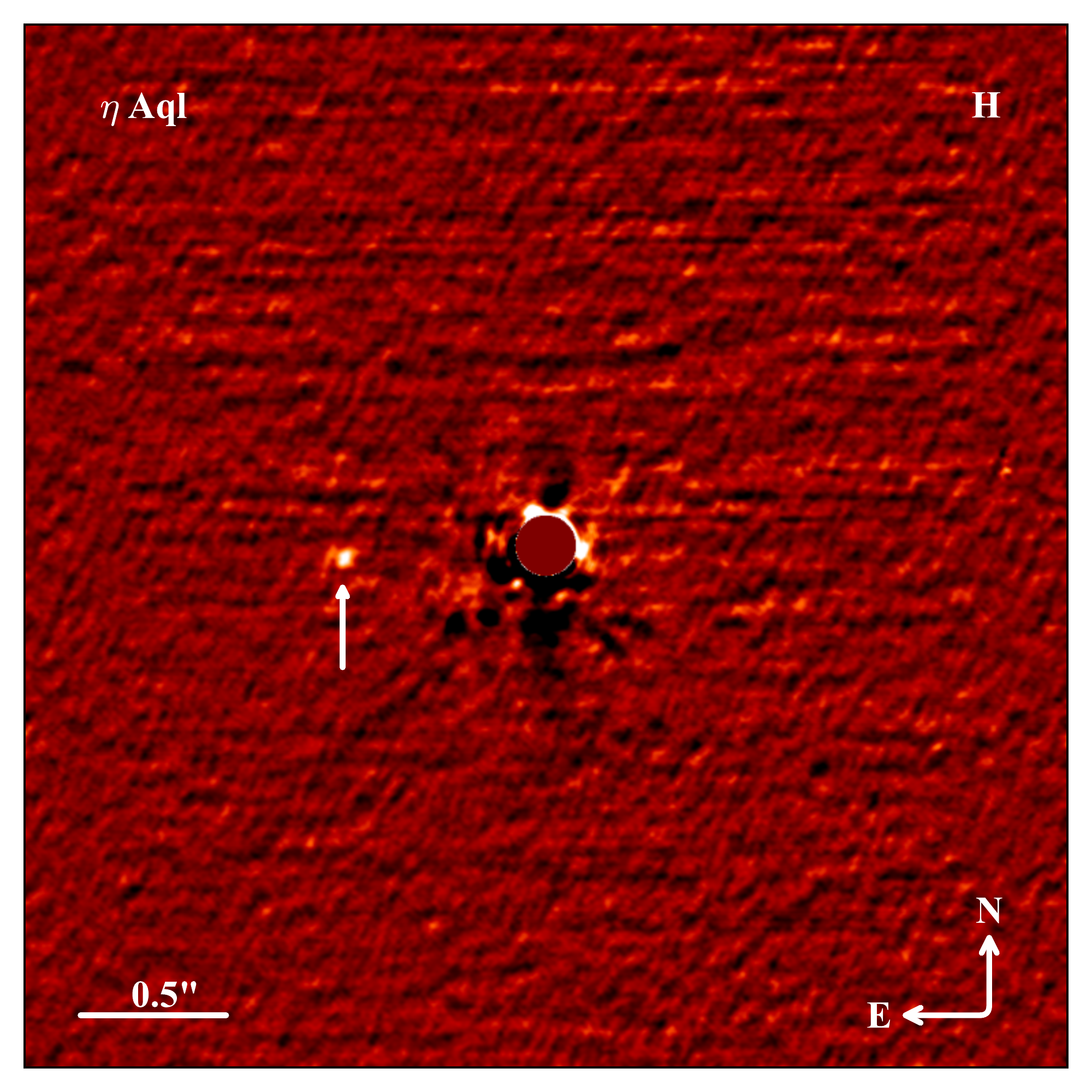}
\includegraphics{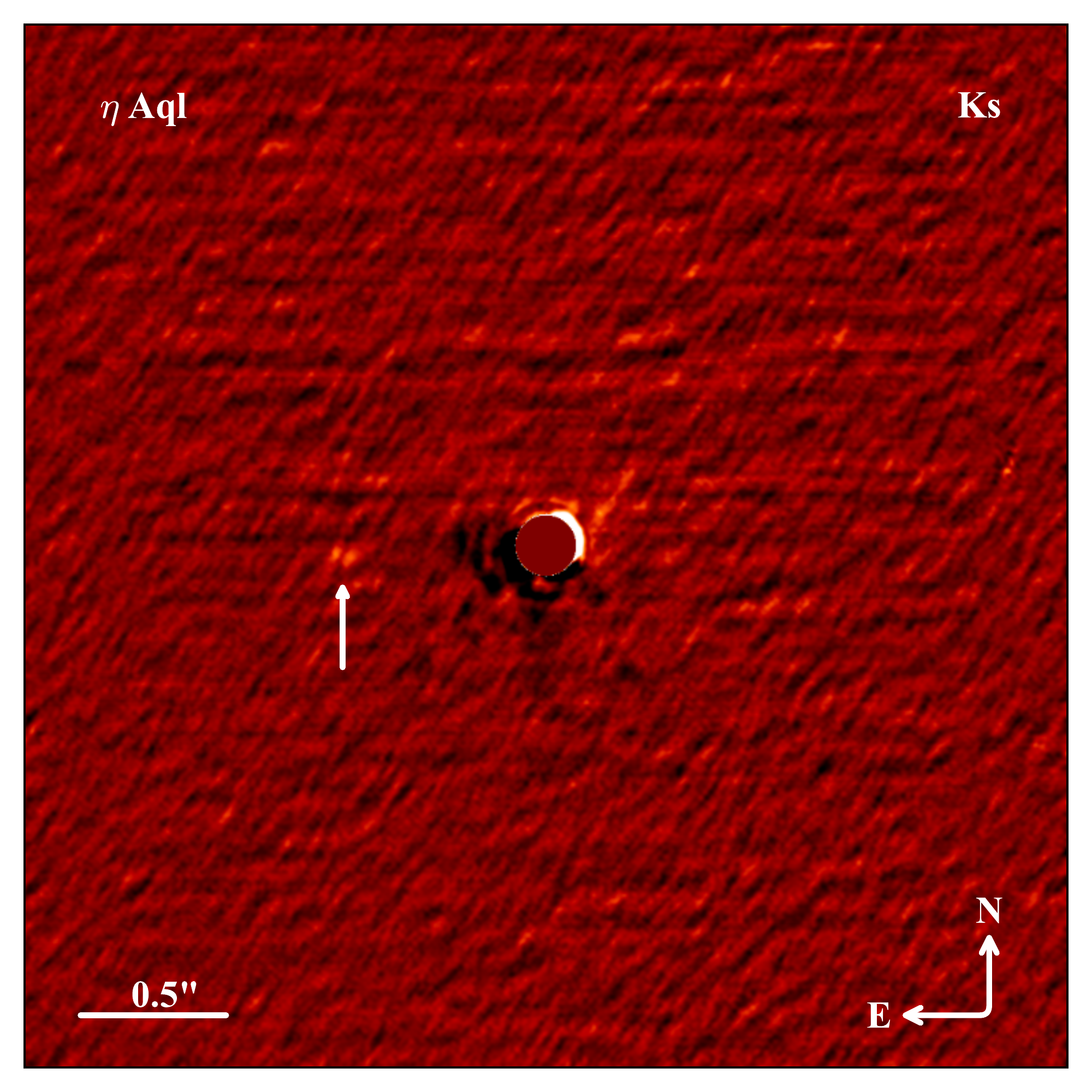}}
\resizebox{\hsize}{!}{\includegraphics{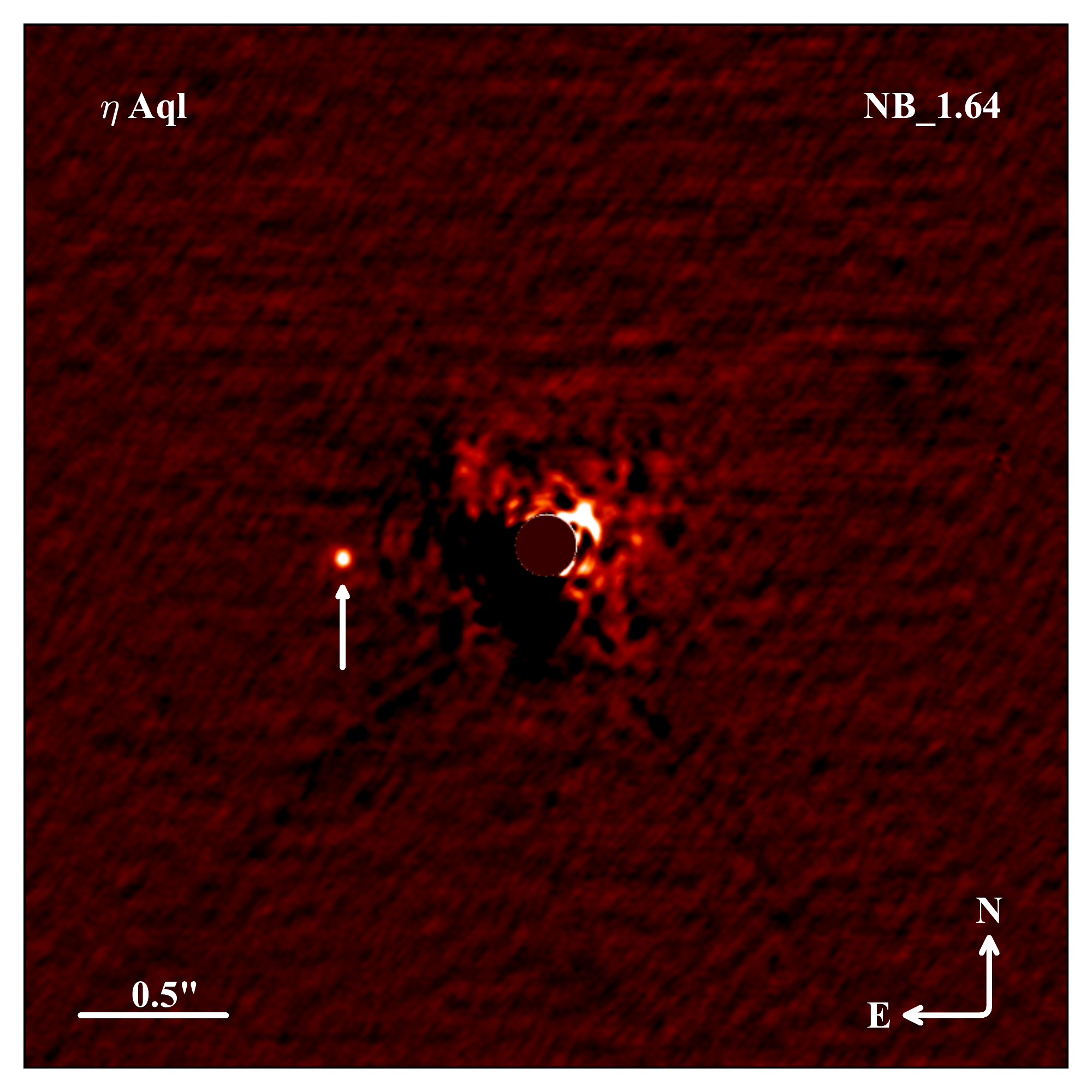}
\includegraphics{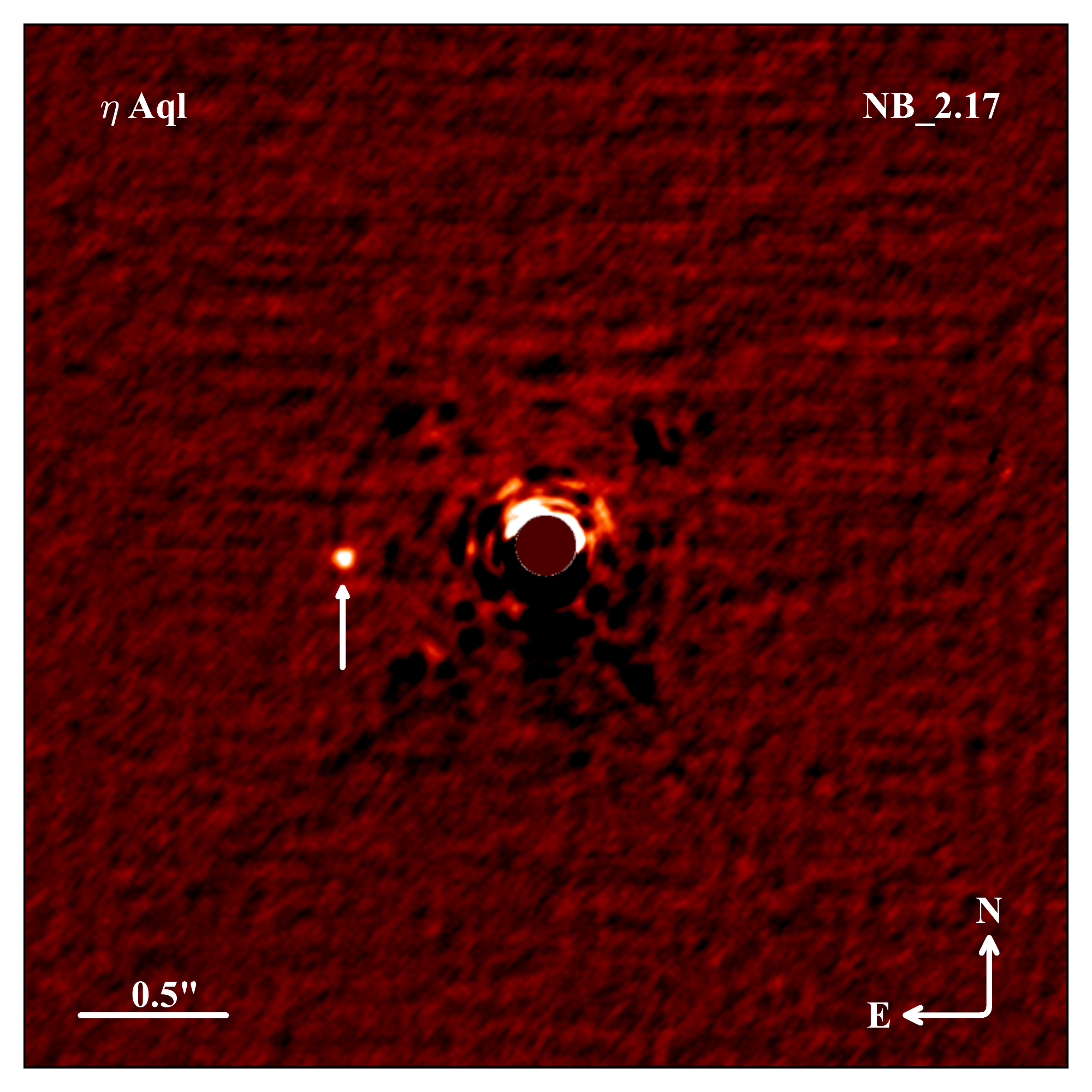}}
\caption{PSF-subtracted images of $\eta$~Aql. The position of the companion is indicated with the white arrow.}
\label{image__subtracted_image}
\end{figure*}

The companion orbiting $\eta$~Aql is clearly detected in the PSF-subtracted images shown in Fig.~\ref{image__subtracted_image}, although some artefacts (spider's pattern, etc.) are still present due to an imperfect PSF subtraction. Because of the neutral density filter, the companion appears brighter in the narrow band filters.

For the other Cepheids, no companion is detected, which might mean that no wide component is present or it is too faint in these bands. We can, however, set upper limits on the magnitude difference for our observing wavelengths.

\begin{figure*}[!t]
\centering
\resizebox{1.\hsize}{!}{
\includegraphics{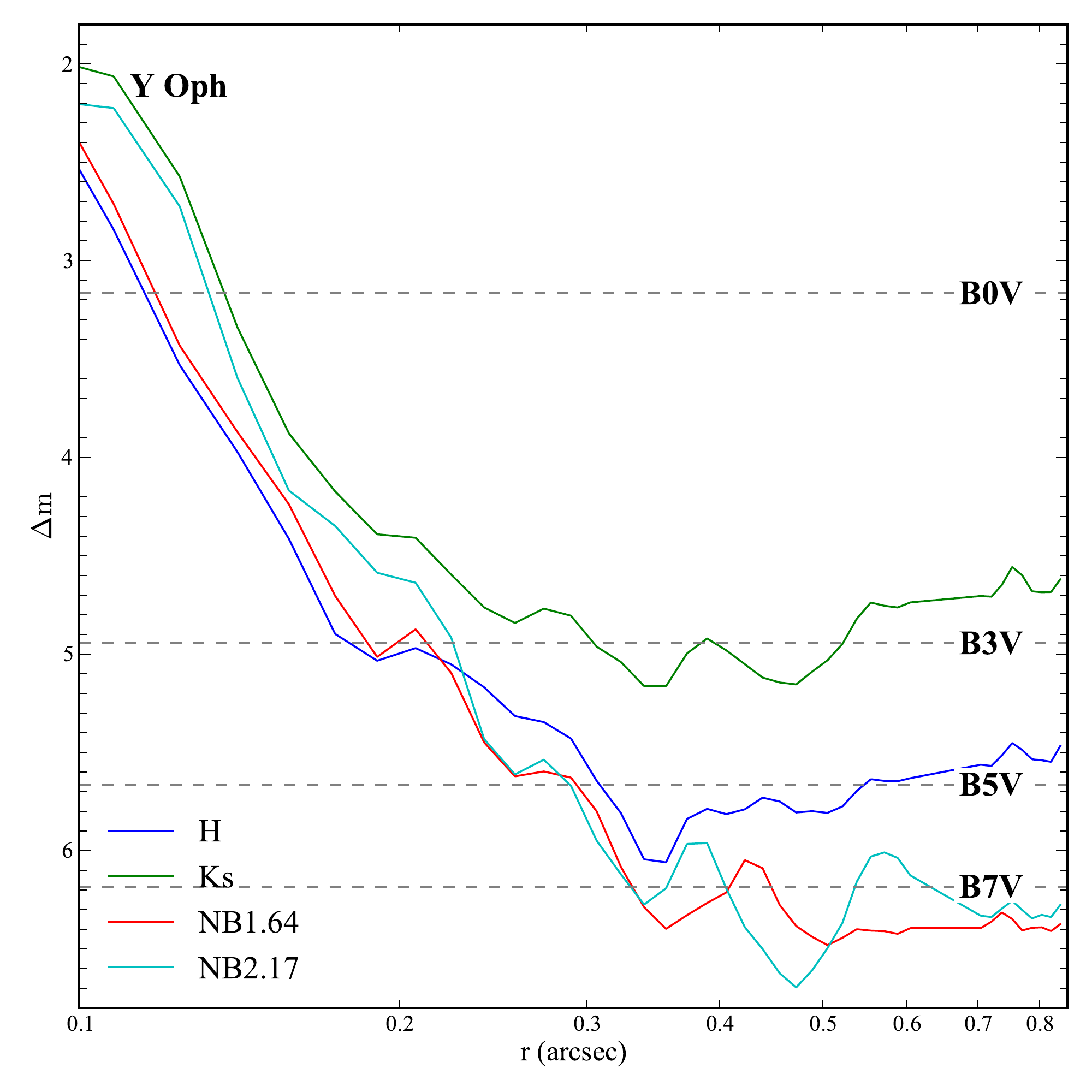}\vspace{.05cm}
\includegraphics{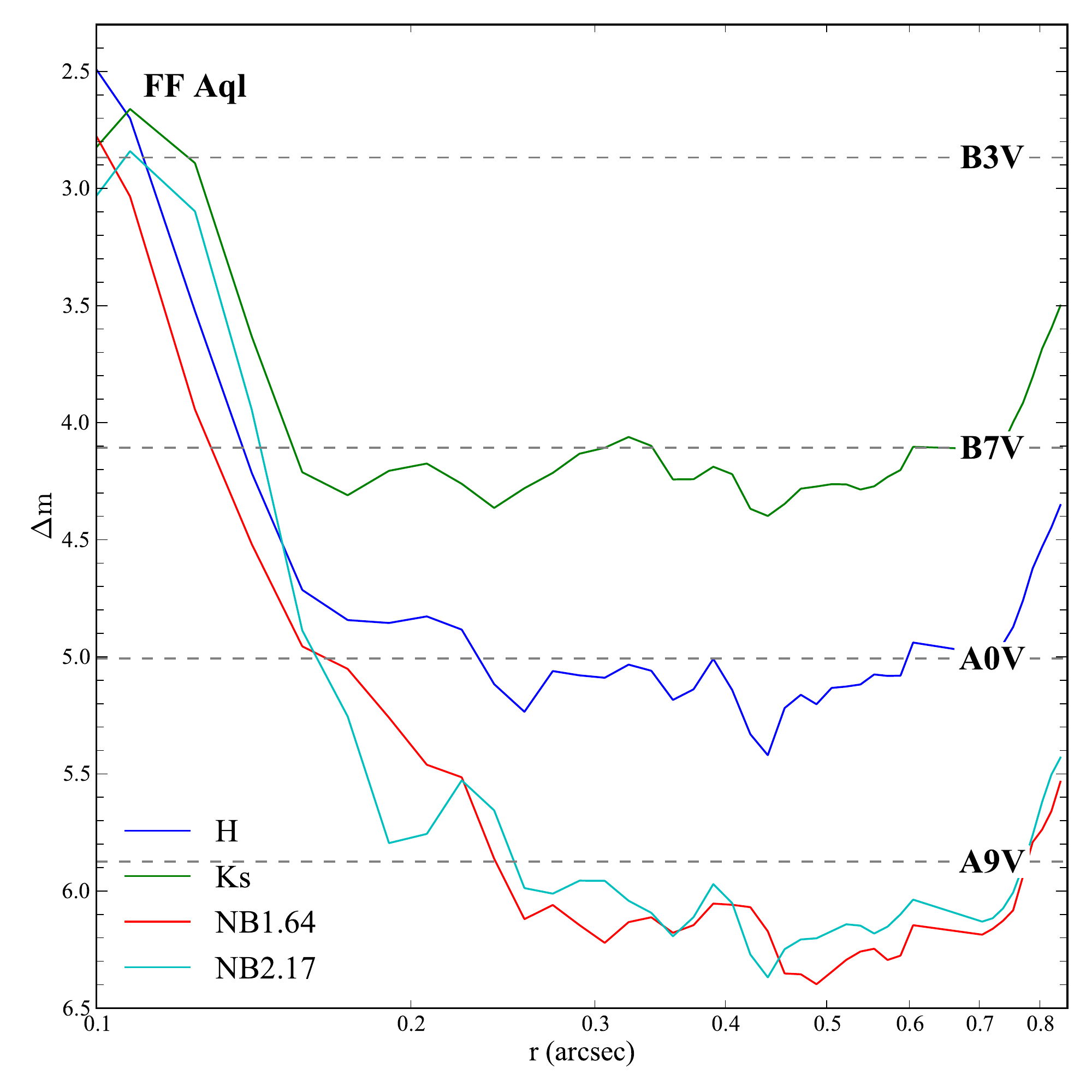}\vspace{.05cm}
\includegraphics{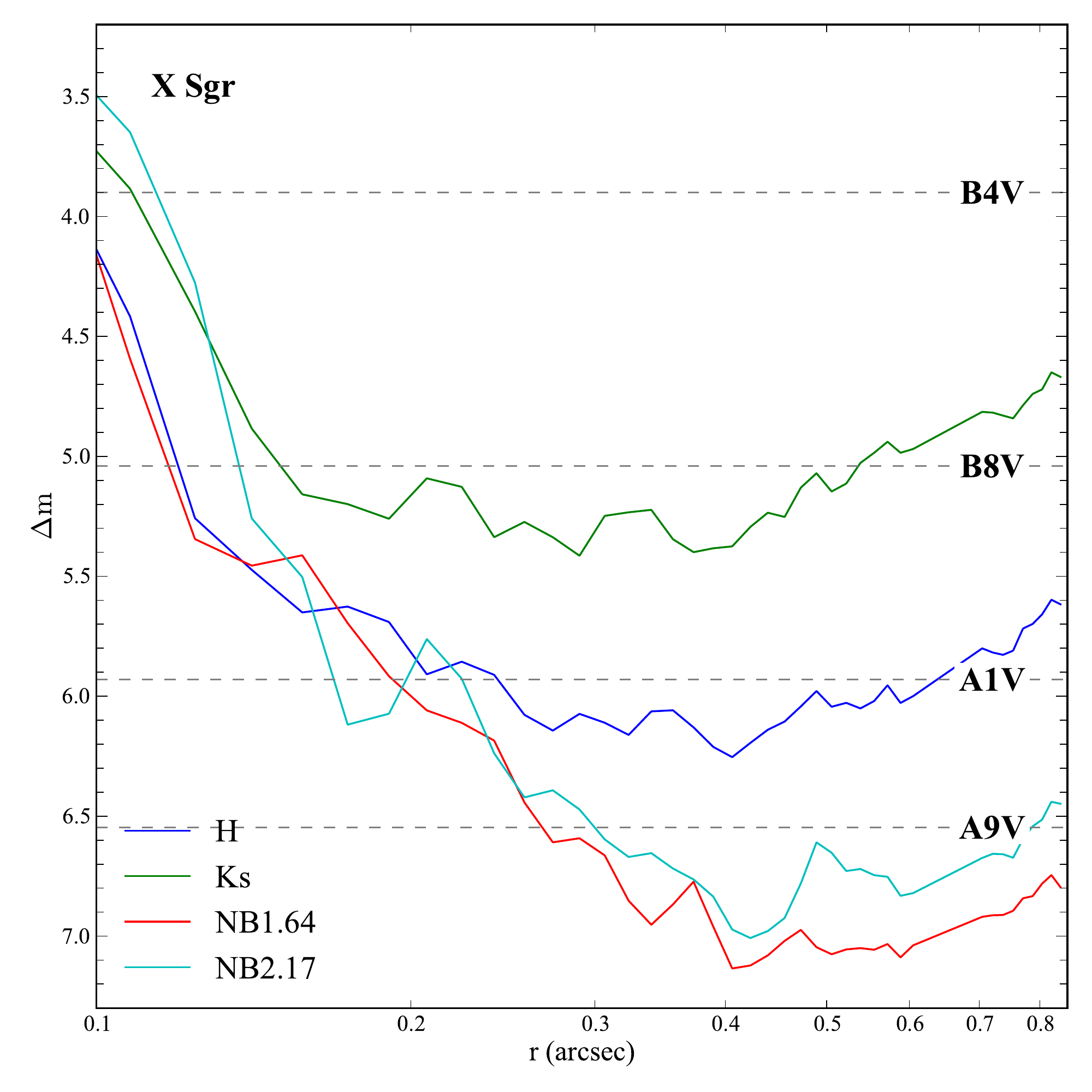}}
\resizebox{.67\hsize}{!}{
\includegraphics{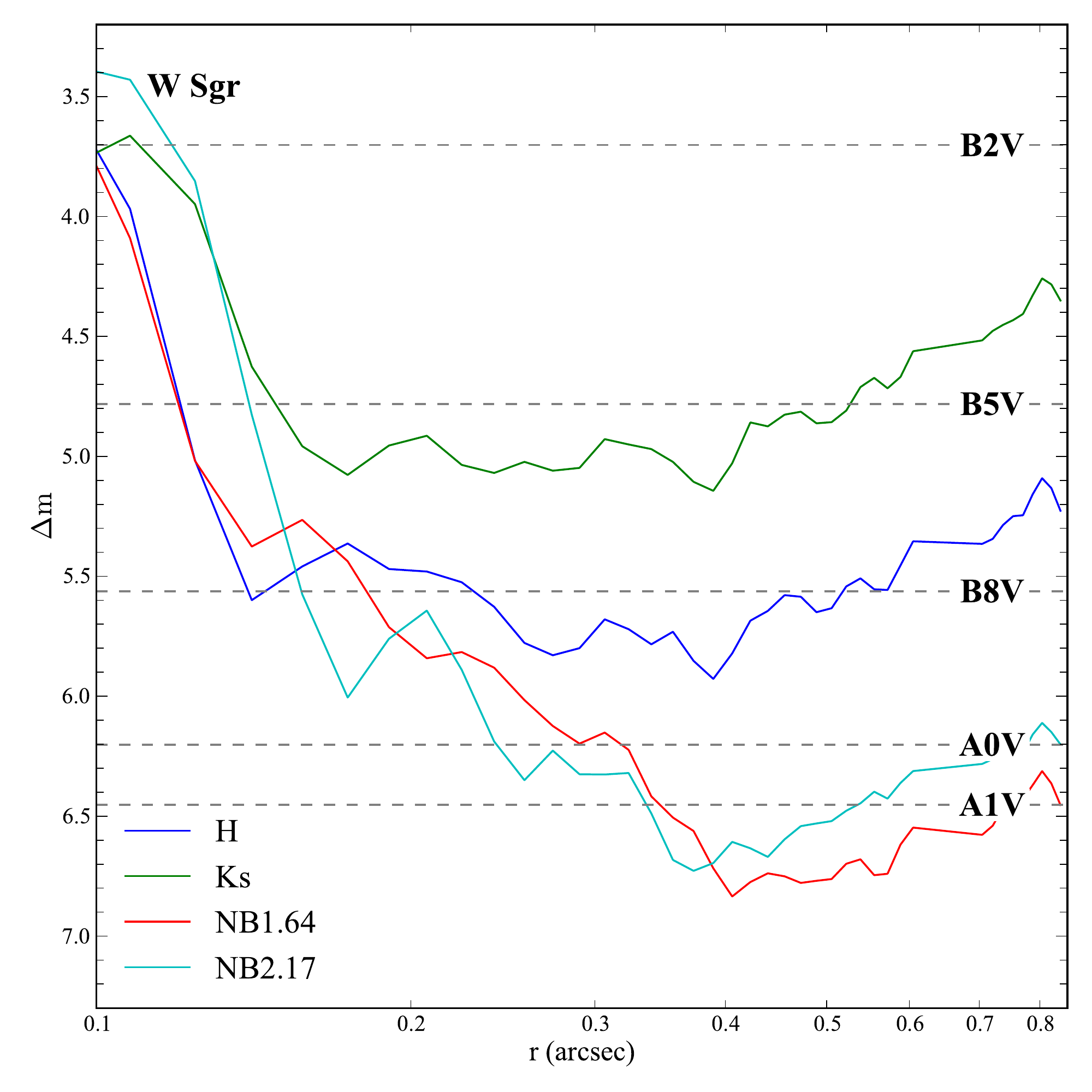}\vspace{.05cm}
\includegraphics{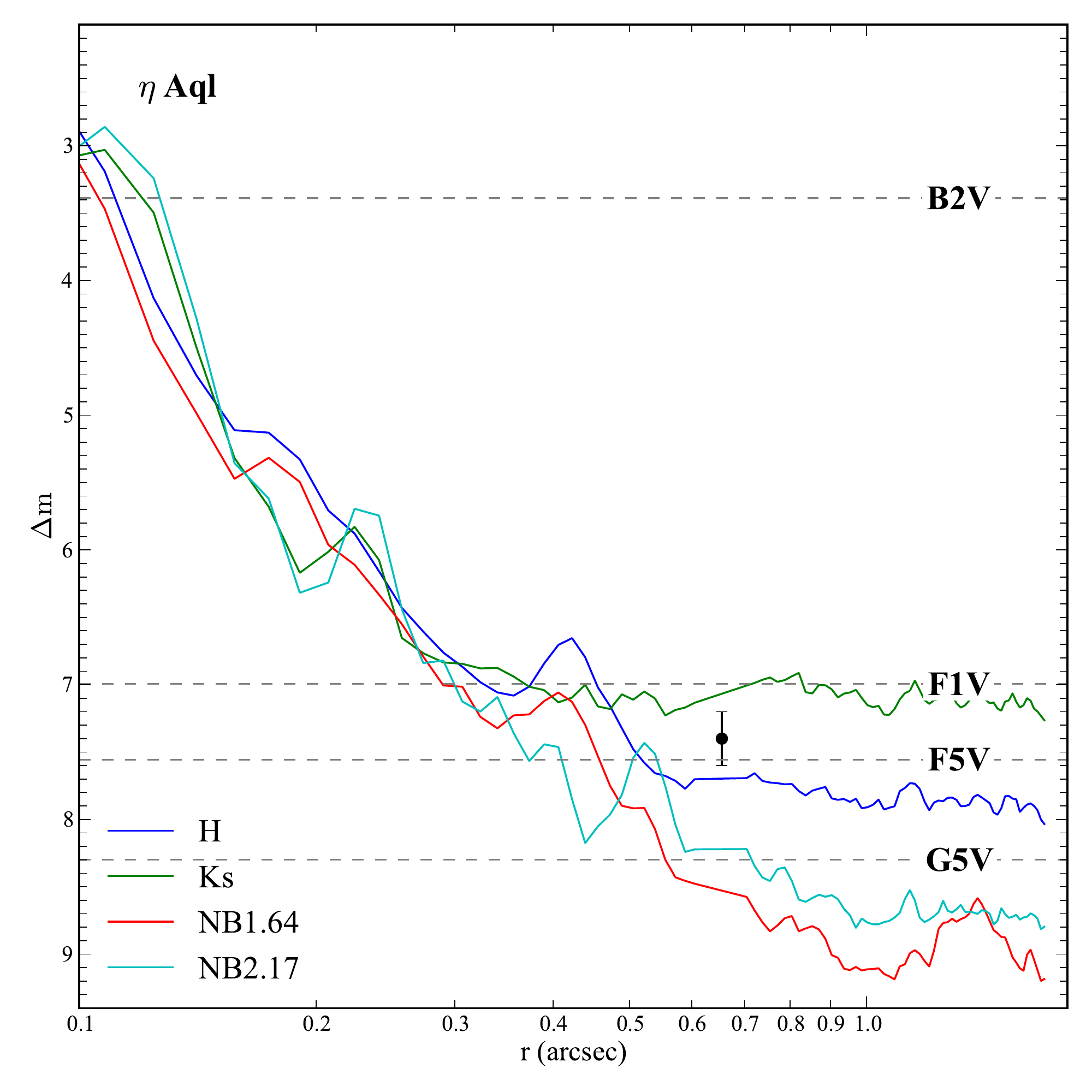}}
\caption{VLT/NACO $5\sigma$ detection limits as a function of angular distance for Y~Oph, FF~Aql, X~Sgr, W~Sgr and $\eta$~Aql, in all filters. The dot in the $\eta$~Aql plot denotes the measured contrast of the detected companion in the $H$ band.}
\label{image__subtracted_image1}
\end{figure*}

\section{Detection limits}
\label{section__detection_limits}

For each star, we estimated the $3\sigma$ detection limit by evaluating the noise within azimuthal rings divided by the primary star maximum flux \citep{Chauvin_2012_12_0, Masciadri_2005_06_0}. The curves are reported in Fig.~\ref{image__subtracted_image1} for all filters. To convert from pixels to angular values, we used the pixel scale $13.26 \pm 0.03$\,mas/pixel \citep{Masciadri_2003_11_0}.

To set an average detection limit, we estimated the radial mean for $r > 0.2\arcsec$ for each filter, and we used the standard deviation as uncertainty. We also interpolated with a periodic cubic spline function the $H$ and $K$ band light curves available in the literature to estimate the Cepheid magnitude at our given phase and determine the companion magnitude limits.

\subsection{Y~Oph}

The binary nature of this Cepheid was first detected by \citet{Abt_1978_04_0} from radial velocity (RV) measurements with an orbital period of 2612\,days. \citet{Pel_1978_01_0} found that the star is too blue for its pulsation period in the ($V - B$) color and suspected a possible blue companion. However, \citet{Evans_1992_01_0} did not detect it in the IUE spectra and set an upper limit for the companion spectral type of A0V star. \citet{Szabados_1989_01_0} gathered all RVs and revised the orbital period to 1222.5\,days. The companion was also not detected by \citet{Hartkopf_1984_01_0} from speckle interferometry, leading to a magnitude difference $\Delta m_\mathrm{V} > 2$ for projected separations $> 30$\,mas.

From our NACO images, no companions are detected. Figure~\ref{image__subtracted_image1} shows that the companion has to be cooler than a B6V star if it is located at $r > 0.2\arcsec$. We derived the contrast limits, $\Delta H > 5.6 \pm 0.2$\,mag and $\Delta Ks > 4.8 \pm 0.2$\,mag for $r > 0.2\arcsec$. In the narrow-band filters, we estimated a dynamic range higher than $6.1 \pm 0.2$\,mag for the same radius range. Using the $H$ and $K$ magnitudes from \citet{Laney_1992_04_0}, this translates to a magnitude limit for the companion of $m_\mathrm{H} > 8.4 \pm 0.2$\,mag and $m_\mathrm{Ks} > 7.5 \pm 0.5$\,mag. For $0.1\arcsec < r < 0.2\arcsec$, we can safely derive $\Delta H \backsimeq \Delta K > 2.5$\,mag, which leads to $m_\mathrm{H} > 5.3$\,mag and $m_\mathrm{Ks} > 5.1$\,mag.

\subsection{FF~Aql} This is a possible quadruple system with a visual component. The spectroscopic companion was first detected by \citet{Abt_1959_11_0} from RV measurements, although the total amplitude of the $\gamma$-velocity is small. \citet{Szabados_1977_01_0} also explained the scatter in the O-C RV diagram as due to orbital motion. \citet{Evans_1990_05_0} performed a complete study of this system and deduced the orbital elements by combining new and old RV measurements to derive an orbital period of 1430\,days. \citet{Benedict_2007_04_0} included orbital perturbations to their HST parallax fit and derived all the orbital elements of the spectroscopic system (semimajor axis of 12.8\,mas).

Wider companions are, however, more uncertain. \citet{McAlister_1987_03_0,McAlister_1989_02_0} resolved a companion at two different epoch, but with two inconsistent separations (0.15\arcsec\  and 0.23\arcsec) for a same position angle (PA). There are still no proofs about a physical association. \citet{Jeffers_1963__0} reported the detection of an additional visual component at separations ranging 6.3\arcsec-6.8\arcsec\ with PA from 132\degr\ to 141\degr\ (from 1886 to 1959). Recently from adaptive optics imaging, however,\citet{Roberts_2011_05_0} also detected the component at 6.8\arcsec\ with a similar PA (144\degr in 2002). There are three possible explanations. First, this system might have an orbital period of about 43\,yr ; however, this would not be consistent with the minimum period. Indeed, by using Kepler’s law of a triple system (Cepheid+spectroscopic+wide companion, we supposed the speckle one does not exist), assuming the measured projected separation 6.8\arcsec\ is a lower limit for the angular semimajor axis, a distance $d = 356$\,pc \citep{Benedict_2007_04_0}, and taking a total mass of about $10\,M_\odot$ (assuming a mass ratio of unity for each component with respect to the Cepheid and with 3.2\,$M_\odot$ for FF Aql, see Sect.~\ref{section__discussion}), the minimal orbital period would be $\sim 100$\,yr. The second explanation would be that the component did not move from 6.8\arcsec, meaning that the orbital period would be longer than 43\,yr, which is even longer than the period limit. The last explanation would be that this star did not move because it is not physically bound to the system. This was also the conclusion of \citet{Evans_1990_05_0}, who detected a companion with a spectral type between A9V and F3V from IUE spectra and concluded that it is the spectroscopic one.

In our observations, the visual component is out of the NACO FoV ; the spectroscopic one is inside the PSF and cannot be detected, and we did not detect the speckle companion. For $r > 0.2\arcsec$, we derived average upper limits of $\Delta H > 5.0 \pm 0.2$\,mag and $\Delta Ks > 4.1 \pm 0.2$\,mag, which convert to $m_\mathrm{H} > 8.6 \pm 0.2$\,mag and $m_\mathrm{Ks} > 7.6 \pm 0.2$\,mag, using the light curves from \citet{Welch_1984_04_0}. For $NB1.64$ and $NB2.17$, we estimated a magnitude difference larger than $6.0 \pm 0.2$\,mag. For $0.1\arcsec < r < 0.2\arcsec$, we conservatively estimated $\Delta H \backsimeq \Delta K > 2.8$\,mag, which leads to $m_\mathrm{H} > 6.4$\,mag and $m_\mathrm{Ks} > 6.3$\,mag.

\subsection{X~Sgr}
A large scatter in the RV data was observed by \citet{Lloyd-Evans_1968__0}, but he did not suspect a possible orbital effect caused by a companion. \citet{Szabados_1990_01_0} found that the velocity curve is best fitted by including an orbital period and derived a period of about 507\,days. The IUE observations were also carried out on this star by \citet{Evans_1992_01_0}, but they did not detect the companion. They, however, set an upper limit for the spectral type of A0V. The binary nature was also analyzed by \citet{Groenewegen_2008_09_0} and derived a slightly longer orbital period of about 574\,days.

More recently, using the AMBER interferometric recombiner, \citet{Li-Causi_2013_01_0} reported the detection of the companion at a projected separation of 10.7\,mas with a flux ratio of about 0.6\,\% ; however, such a result has to be confirmed as this detection is at the capability limits of the instrument \citep[see e.g.][]{Absil_2010_09_0}



The spectroscopic component is not detectable from our NACO observations, while other wide companions are not detected. We set contrast limits for a wide component at $r > 0.2\arcsec$ of $\Delta H > 6.0 \pm 0.2$\,mag and $\Delta Ks > 5.0 \pm 0.2$\,mag, transforming $m_\mathrm{H} > 8.7 \pm 0.2$\,mag and $m_\mathrm{Ks} > 7.7 \pm 0.2$\,mag \citep[using the light curves from][]{Feast_2008_06_0}. We also measured a mean dynamic range higher than $6.6 \pm 0.2$\,mag in the narrow band filters. For $0.1\arcsec < r < 0.2\arcsec$, we estimated $\Delta H \backsimeq \Delta K > 3.9$\,mag, leading to $m_\mathrm{H} > 6.7$\,mag and $m_\mathrm{Ks} > 6.5$\,mag.

\subsection{W~Sgr}

This is a triple system composed of a spectroscopic and a visual component. The multiplicity of this star has been studied by \citet{Babel_1989_06_0} who derived the first spectroscopic orbital elements of the system, with an orbital period of 1780\,days. \citet{Benedict_2007_04_0} later deduced all the orbital elements from HST observations, including a semimajor axis of 12.9\,mas, and a revised period of 1582\,days. \citet{Evans_2009_03_0} constrained the spectral type of this close companion to be later than an F0V star.

The visual companion was first resolved by \citet{Morgan_1978_06_0} at $\sim 0.12\arcsec$ from speckle interferometry ; however, it was not detected by \citet{Bonneau_1980_11_0} with the same observing technique. Then, a hot companion is detected from IUE spectra \citep{Bohm-Vitense_1985_09_0, Evans_1991_05_0} with a spectral type A0V, which might be the wide component. \citet{Evans_2009_03_0} confirmed these results with the detection of both companions with HST STIS with the widest star located at $\sim 0.16\arcsec$.

This wide component is inside the NACO FoV ; however, it is just below our detection limits. We estimated a contrast to be $\Delta H > 5.5 \pm 0.2$\,mag and $\Delta Ks > 4.7 \pm 0.3$\,mag for $r > 0.2\arcsec$, which convert to $m_\mathrm{H} > 8.4 \pm 0.2$\,mag and $m_\mathrm{Ks} > 7.5 \pm 0.3$\,mag \citep[using light curves of][]{Welch_1984_04_0}. We found a contrast higher than $6.4 \pm 0.3$\,mag in the narrow-band filters. For $0.1\arcsec < r < 0.2\arcsec$, we estimated $\Delta H \backsimeq \Delta K > 3.7$\,mag, leading to $m_\mathrm{H} > 6.6$\,mag and $m_\mathrm{Ks} > 6.4$\,mag. 

\subsection{$\eta$~Aql}

The first evidence of a companion was given from IUE spectra \citep{Mariska_1980_06_0}, where a significant UV emission corresponding to an A0V star is detected. However, there are no signs of short-term orbital effects in the RV measurements, which would be consistent with a wide component with low velocity amplitude. Additional IUE spectra later confirmed the presence of a hot companion near $\eta$~Aql \citep{Evans_1991_05_0} and refined the spectral type to be B9.8V. However, this component was not detected from speckle interferometry \citep{Mason_1999_04_0} with a typical resolution limit of 40\,mas. No detection was also reported from the Hipparcos telescope \citep{Mason_1999_04_0}, which set typical detection limits in the $V$ band of $\Delta m = 0.8, 2.7$, and 4.0\,mag for separation $\rho < 0.1, 0.2$, and 0.5\arcsec, respectively. From HST FGS observations, \citet{Benedict_2007_04_0} found some astrometric perturbations in their measurements, but they were not able to fit an orbital motion because of a lack of constraints from radial velocity variations. \citet{Remage-Evans_2013_10_0} resolved a companion at 0.66\arcsec\ with the HST telescope. This companion is located too far from the Cepheid to be the one detected by \citet{Benedict_2007_04_0}, and as mentioned by \citet{Remage-Evans_2013_10_0}, this would indicate the presence of a third closest component.

A companion is detected in our NACO images, and its position is consistent with the one resolved with HST. The $\eta$~Aql system is analyzed more particularly in Sect.~\ref{section__the_case_of_eta_aql}. We also derived upper magnitude limits to exclude the presence of other companions in our FoV. For $r > 0.2\arcsec$, we estimated $\Delta H > 7.6 \pm 0.5$\,mag and $\Delta Ks > 7.0 \pm 0.2$\,mag, which convert to $m_\mathrm{H} > 9.7 \pm 0.5$\,mag and $m_\mathrm{Ks} > 9.0 \pm 0.2$\,mag \citep[using light curves from][]{Barnes_1997_06_0}. For $NB1.64$ and $NB2.17$, we estimated a magnitude difference larger than $8.3 \pm 0.2$\,mag. In the range $0.1\arcsec < r < 0.2\arcsec$, we evaluated $\Delta H \backsimeq \Delta K > 3.4$\,mag, giving $m_\mathrm{H} \sim m_\mathrm{Ks} > 5.4$\,mag.

\section{The wide component of $\eta$~Aql}
\label{section__the_case_of_eta_aql}

The wide component around $\eta$~Aql is clearly detected in the PSF-subtracted images of Fig.~\ref{image__subtracted_image}, except in the $Ks$ band where its detection is marginal. In this section, we estimate its angular position relative to the Cepheid and its magnitude in the corresponding filters. It is worth mentioning, however, that its physical association to the Cepheid has not been proven yet.

\subsection{Astrometric position}

\begin{figure}
\centering
\resizebox{\hsize}{!}{\includegraphics{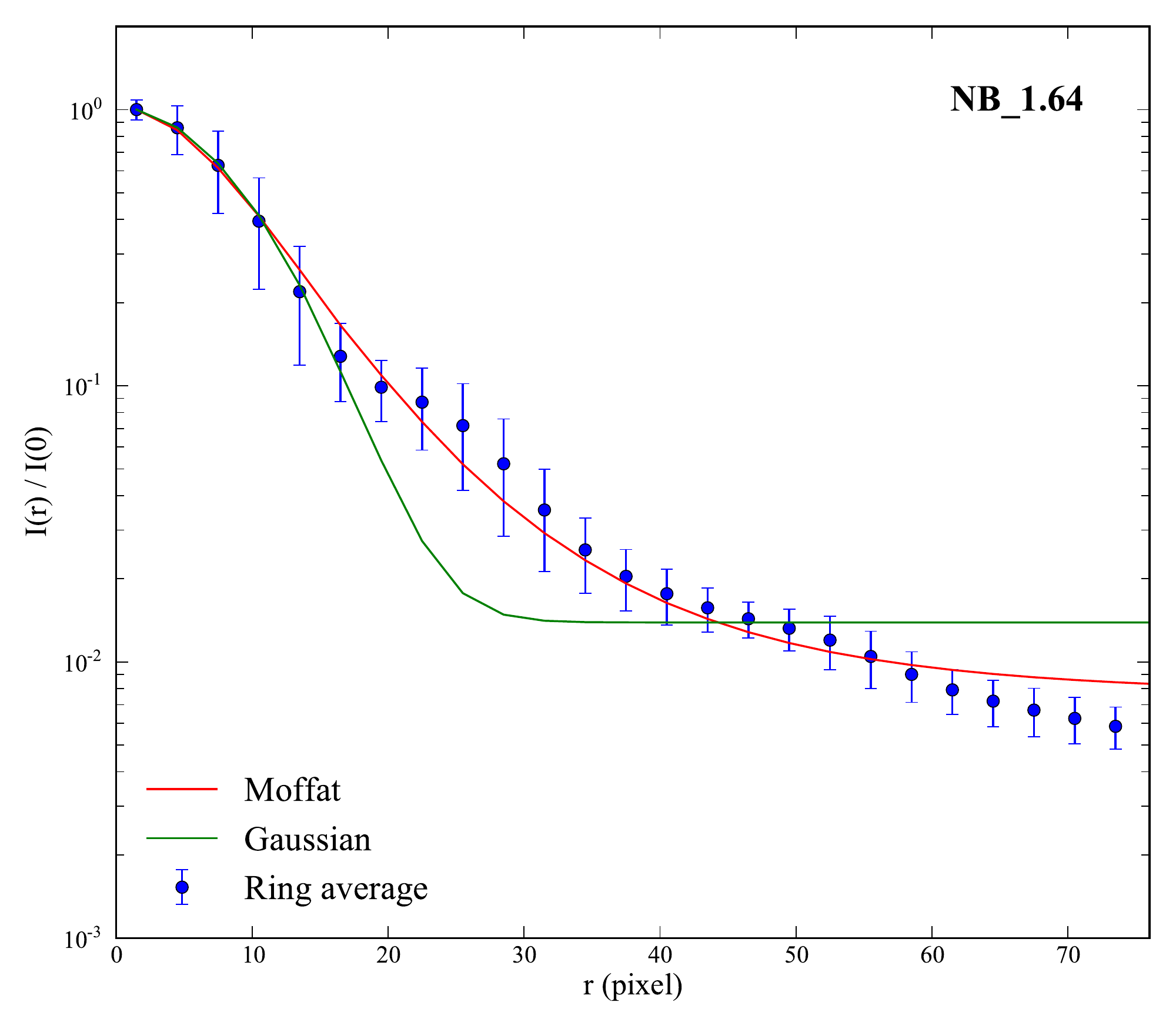}\includegraphics{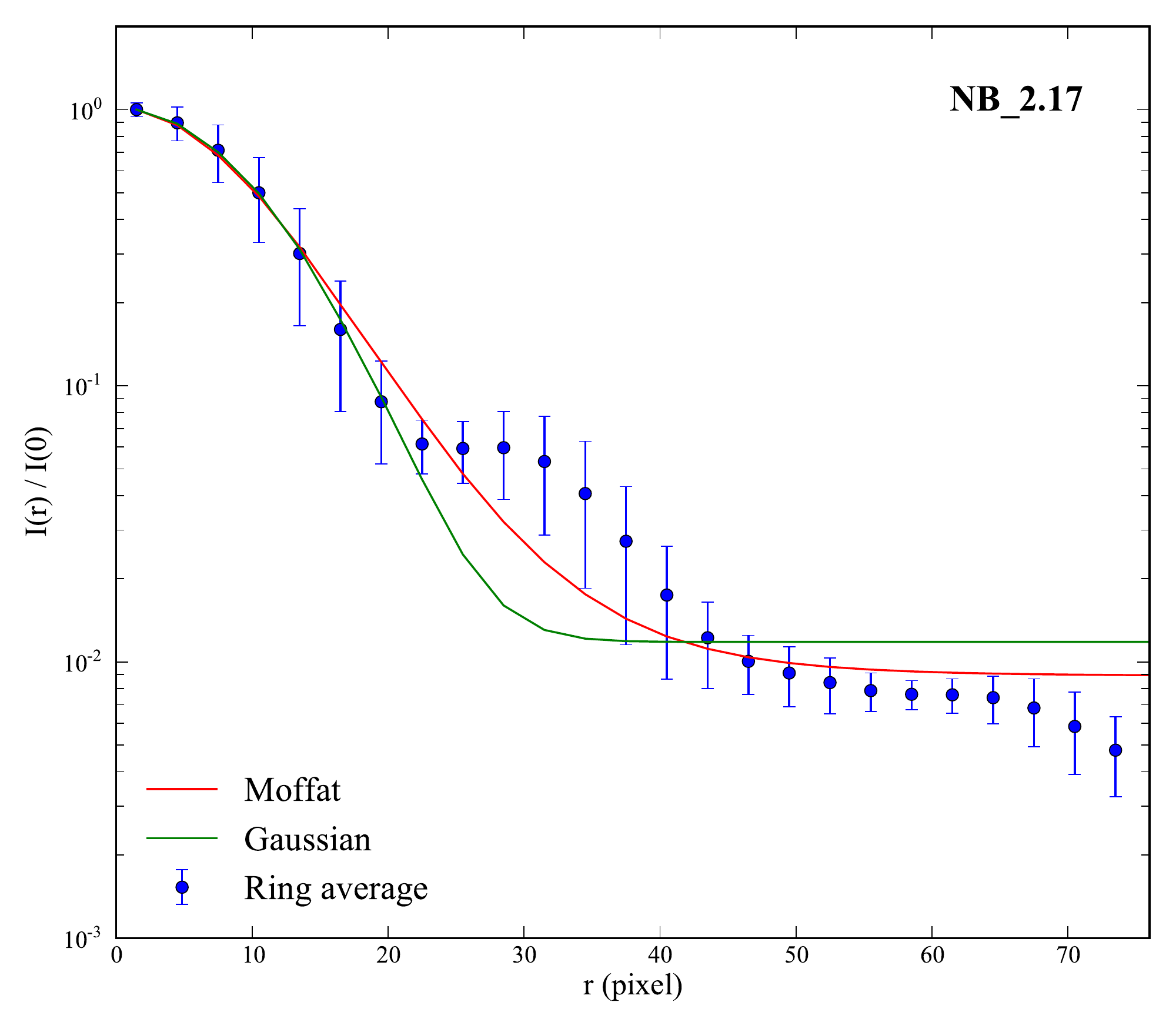}}
\caption{Radial profile of the $\eta$~Aql images in the narrow band filters. Radial flux was evaluated estimating the average value and standard deviation in rings.}
\label{image__moffat}
\end{figure}

We estimated the relative position by fitting a 2D-Moffat and 2D-Gaussian function, respectively for the Cepheid and the companion, which have the following form
\begin{eqnarray*}
I(x, y)_\mathrm{ceph} &=& I_\mathrm{0,ceph} \left[ \left( \frac{x-x_0}{\rho_x} \right)^2  + \left( \frac{y-y_0}{\rho_y} \right)^2 + 1 \right]^{-\beta} + c_1,\\
I(x, y)_\mathrm{comp} &=& I_\mathrm{0,comp} \exp \left( -\frac{(x-x_0)^2}{2\sigma_x^2}  - \frac{(y-y_0)^2}{2\sigma_y^2} \right) + c_2,
\end{eqnarray*}
where $x_0, y_0$ are the coordinates of the center of each star, ($\rho_x, \rho_y$) denote the full width at half maximum of the PSF, ($\sigma_x, \sigma_y$) the standard deviations of the Gaussian, ($c_1, c_2$) are constants, and  $\beta$ is a variable power law index. We chose a 2D-Moffat function for the Cepheid, because it has the advantage to better fit the AO PSF wings, as shown in Fig~\ref{image__moffat}. Indeed, the PSF wings are more extended because of the residual light out of the coherent core (AO halo) for bright stars.

We extracted subwindows of $76\times76$ pixels ($0.25\arcsec\times0.25\arcsec$) for the Cepheid and  $30\times30$ pixels ($0.10\arcsec\times0.10\arcsec$) for the companion. We measured the relative position in the narrow band filters NB\_2.17 and NB\_1.64, in which the signal-to-noise ratio of the companion is higher. In the broadband filters, the flux of the companion is too low to obtain a consistent fit. Uncertainties were estimated using the bootstrapping technique with 500 bootstrap samples. Enlarging the sub-window has an impact of at most 0.5\,\% on $\Delta \delta$, which is well below our accuracy level, while it is negligible on $\Delta \alpha$.

Refraction effects by the Earth's atmosphere causes an apparent change in the true astrometric position. These effects depend on the wavelength and the atmospheric column depth. In our case, the chromatic differential refraction has no impact as we used narrow-band filters. However, the achromatic refraction has to be corrected for. We estimated these corrections, following the method of \citet{Kervella_2013_04_0}, which uses the \texttt{slarefro} function distributed with the \texttt{Starlink} library\footnote{We actually used the \texttt{pySLALIB} module which contains Python wrappers for every Fortran functions in the SLALIB library.}. The right ascension and declination corrections to add to the measured differential astrometry are the values, $\Delta \alpha_\mathrm{corr} =  171.42\,\mathrm{\mu as}$ and $\Delta \delta_\mathrm{corr} = -130.04\,\mathrm{\mu as}$ at $2.166\,\mathrm{\mu m}$, and $\Delta \alpha_\mathrm{corr} = 170.94\,\mathrm{\mu as}$ and $\Delta \delta_\mathrm{corr} = -132.41\,\mathrm{\mu as}$ at $1.644\,\mathrm{\mu m}$ (it is worth mentioning that these values are below our accuracy level, i.e. $< 0.15\sigma$). The final astrometric positions are listed in Table~\ref{table__astrometry} with a weighted average value. The values for both filters agree within $1\sigma$.

The companion is located at an average angular separation $\rho = 654.7 \pm 0.9$\,mas and a position angle $PA = 92.8 \pm 0.1\degr$. These values are consistent with the detection of \citet[][and private communication]{Remage-Evans_2013_10_0} using the Hubble Space Telescope.

\begin{table}
\centering
\caption{Relative astrometric position of the companion of $\eta$~Aql, corrected for atmospheric refraction.}
\begin{tabular}{cccc} 
 \hline
 \hline
Filter		&	MJD							& 	$\Delta \alpha$	&	$\Delta \delta$		  \\
			&								&		(mas)			&			(mas)							  \\
 \hline
NB\_1.64	& 	56128.2986		& 		$654.70 \pm 1.16$ & $-32.37 \pm 1.18$	   \\
NB\_2.17	& 	56128.3012		& 		$652.93 \pm 1.24$ & $-30.96 \pm 1.76$	 \\
 \hline
 Avg.   	&	56128.2999		&		$653.87 \pm 0.85$ & $-31.93 \pm 0.98$	\\
 \hline
\end{tabular}
\label{table__astrometry}
\end{table}

\subsection{Photometry}

There is no existing photometric template for HD~188512 ; we, therefore, created a synthetic spectral energy distribution (SED) following the method used by \citet{Cohen_1999_04_0} and \citet{Merand_2005_04_0}. We fitted stellar atmosphere models obtained with the ATLAS9 simulation code from \citet{Castelli_2003__0} for the wavelength range $0.4-25\,\mathrm{\mu m}$. We chose a grid that was computed for solar metallicity and a turbulence velocity of $2\,\mathrm{km\,s^{-1}}$. We then interpolated this grid to compute spectra for any effective temperature and any surface gravity. The spectrum was multiplied by the solid angle of the stellar photosphere, $\pi \theta_{LD}/4$, where $\theta_{LD}$ is the limb-darkened angular diameter. 

 We adjusted the photometric data to the model by taking the spectral response of each instrument into account. We did not adjust the surface gravity, since the broadband photometry is mostly insensitive to this parameter. Therefore, we took an average effective gravity of $\log g = 2.0$. 
 Changing this value by $\pm 1.0$ leads to a variation in the derived parameters of less than 2\,\%, which we also considered in the final uncertainties for the corresponding stars. During the fitting procedure, we also fitted the color excess $E(B - V)$ for all flux densities $<3\,\mathrm{\mu m}$ by adopting the reddening law from \citet{Fouque_2007_12_0}. Data for $\lambda > 3\,\mathrm{\mu m}$ were not corrected for the interstellar extinction, which we assumed to be negligible.

We used photometry from the Tycho-2 Catalogue \citep{Hog_2000_03_0}, the Two Micron All Sky Survey \citep[2MASS,][]{Cutri_2003_03_0}, the Wide-field Infrared Survey Explorer \citep[WISE,][]{Wright_2010_12_0}, the Infrared Astronomical Satellite \citep[IRAS,][]{_1988__0}, and the AKARI satellite IRC point source catalog \citep{Ishihara_2010_05_0}.

We then carried out a classical aperture photometry to estimate the flux density of both component in each filter. Absolute calibration was done by considering the filter transmission\footnote{Filter transmission profiles are available on the website http://svo2.cab.inta-csic.es/theory/fps/.}. We chose an aperture of 0.55\arcsec\ for the central source (Cepheid and reference star) and a sky annulus thickness of 0.2\arcsec\ from 0.80\arcsec to avoid the companion. For the component photometry, we took an aperture of 0.06\arcsec and a sky annulus of 0.03\arcsec\ thickness with an internal radius of 0.09\arcsec. 

We determined dereddened magnitude by adopting the reddening law from \citet{Fouque_2007_12_0} with a total-to-selective absorption in the $V$ band of $R_\mathrm{V} = 3.23$ \citep{Sandage_2004_09_0} and a color excess $E(B - V) = 0.130$ from \citet{Fouque_2007_12_0}. The dereddened magnitudes are listed in Table~\ref{table__magnitude}. In the $Ks$ band, the flux was too low to obtain a reliable estimate.

\begin{table}
\centering
\caption{Measured dereddened magnitudes.}
 \label{table__magnitude}
\begin{tabular}{ccccc} 
\hline
\hline
 				&	$H_\mathrm{0}$								& $m^\mathrm{1.64}_0$				&	$K\mathrm{s_0}$					&	$m^\mathrm{2.17}_0$	\\
						&	(mag)						&	(mag)						&	(mag)						& (mag)				\\
\hline
$\eta$~Aql	&  		$1.82_{\pm0.04}$		& $1.81_{\pm0.04}$		& $1.80_{\pm0.04}$   &	$1.79_{\pm0.04}$	\\
Companion			&  	$9.34_{\pm0.04}$		& $9.31_{\pm0.04}$ 			 & --  	&	$9.18_{\pm0.04}$	\\
\hline
\end{tabular}
\end{table}


\section{Discussion}
\label{section__discussion}

Using our measured contrast limits in the $H$ and $K$s bands, we can exclude the presence of companions or set an upper limit on their spectral type with respect to the angular distance from the Cepheid. We utilized the absolute magnitudes of the MK classification in \citet{Cox_2000__0} and the IR intrinsic colors of \citet{Ducati_2001_09_0} for main-sequence stars. We also took the distance of the Cepheids (and so the system) from \citet{Benedict_2007_04_0} and \citet{Kervella_2004_03_0}. We can split our analysis in two distance regimes.

\subsection{Case $r > 0.2\arcsec$}

The upper limits for the spectral type (at $3\sigma$) for the undetected companions are shown in Fig.~\ref{image__subtracted_image1}. For Y~Oph, FF~Aql, X~Sgr and W~Sgr in the $H$ band, we can exclude the presence of companions with a spectral type that appear earlier than B5V, A0V, A1V, and B8V, respectively. In the narrow band filters, the spectral-type upper limits are B7V, A9V, A9V, and A1V, respectively. The spectral-type limits for Y~Oph and FF~Aql agree with those derived in the literature \citep[A0V and A1V, respectively, from][see Sect.~\ref{section__detection_limits}]{Evans_1992_01_0}. However, for X~Sgr, the previous limit was an A0V star, we can now exclude companions brighter than A9V stars. No companion later than an A1V star is detected for W~Sgr in this radius range.

For $\eta$~Aql, we can set an upper limit of an F5V star in $H$ and G5V in the narrow band filters. Our observed companion (black dot in Fig.~\ref{image__subtracted_image1}) is detected at $7\sigma$ in the narrow bands, at $3\sigma$ in $H$, and is almost not detected in $K$. We derived a spectral type for this companion ranging between an F1V and F6V star.

We summarized the magnitudes and spectral type limits in Table~\ref{table__limits}.

\begin{table}
\centering
\caption{Magnitude and spectral types limits for the companions for $r > 0.2\arcsec$.}
 \label{table__limits}
\begin{tabular}{ccccccc} 
\hline
\hline
 	Star			&	$\Delta H$	& $m_\mathrm{H}$	&	$\Delta K $	&	$m_\mathrm{K}$	&	$\Delta NB$	&	Sp.~Typ.	\\

\hline
Y~Oph			&	>5.6					&	>8.4			&	>4.8			&	>7.5	&	>6.1	&	B7V	\\
FF~Aql			&	>5.0					&	>8.6			&	>4.1			&	>7.6	&	>6.0	&	A9V	\\
X~Sgr			&	>6.0					&	>8.7			&	>5.0			&	>7.7	&	>6.6	&	A9V	\\
W~Sgr			&	>5.5					&	>8.4			&	>4.7			&	>7.5	&	>6.4	&	A1V	\\
$\eta$~Aql	&	>7.6					&	>9.7			&	>7.0			&	>9.0	&	>8.3	&	G5V	\\
\hline
\end{tabular}
\tablefoot{$\Delta NB$ denotes the contrast limit in the narrow band filters. Sp.~Typ. represents the spectral-type upper limit in the narrow band filters.}
\end{table}

\subsection{Case $0.1\arcsec < r < 0.2\arcsec$}

We can also set an upper limit on the spectral type of possible companions in that radius range. For Y~Oph, FF~Aql, X~Sgr, W~Sgr, and $\eta$~Aql, we did not detect companions with a spectral type that is earlier than an O9V, B3V, B4V, B2V, and B2V star, respectively. The A0V star companion orbiting W~Sgr, which is located at $\sim 0.15-0.20\arcsec$, is just below our $3\sigma$ threshold and is not detected.

In this radius range, we cannot have more constraints, and the detection of fainter companions is limited to the AO halo, which is also the residual light out of the coherent core.


\subsection{Detected component}
The component around $\eta$~Aql is bright and wide enough to be spatially resolve. Its projected separation and position angle are consistent with the component detected by \citet{Remage-Evans_2013_10_0}. However, our estimated spectral type, which is between an F1V and F6V star, does not agree with the B9.8 star detected from UV spectra \citep{Evans_1991_05_0}. On the other hand, \citet{Mason_1999_04_0} did not detect a component earlier than an A1V star by speckle interferometry at such a separation. Therefore, the B9.8 star should be in a closer orbit, and our detected component, if bound to the system, would be another companion. Additional photometric observations in various other bands are needed to fully constrain its spectral type, while more astrometric measurements will provide information about its association with the $\eta$~Aql system.

\section{Conclusion}
\label{section__conclusion}

We presented high angular resolution imaging with VLT/NACO to search for wide components around five classical Cepheids. We detected a component close to $\eta$~Aql at a separation of 0.65\arcsec\ with a spectral type between F1V and F6V. We measured its dereddened apparent magnitude in the infrared to be $H_0 = 9.34$\,mag. No other companion is present in the FoV.

In the range $r > 0.2\arcsec$, we ruled out the presence of companions that come earlier than B7V, A9V, A9V, A1V, and G5V, respectively for Y~Oph, FF~Aql, X~Sgr, W~Sgr and $\eta$~Aql.

The AO imaging is an useful tool to search for Cepheid companions because it reaches high detection limits. Other techniques, such as differential imaging, aperture masking or coronography, also have the capabilities to detect faint companions. We will be able to reach higher contrasts with the next generation instruments, such as VLT/SPHERE or Gemini/GPI.

Studying Cepheid companions is particularly important to understand the evolution of the pulsating star. Binary Cepheids are also powerful tools in estimating the masses and distances with a unique accuracy. Gaia will revolutionize this field by providing a micro-arcsecond astrometric precisions, and will give accurate Cepheid distances and orbital perturbations.


\begin{acknowledgements}
A.G. acknowledges support from FONDECYT grant 3130361. WG an GP gratefully acknowledge financial support for this work from the BASAL Centro de Astrof\'isica y Tecnolog\'ias Afines (CATA) PFB-06/2007. Support from the Polish National Science Center grant MAESTRO 2012/06/A/ST9/00269 and the Polish Ministry of Science grant Ideas Plus (awarded to GP) is also acknowledged. We acknowledge financial support from the “Programme National de Physique Stellaire” (PNPS) of CNRS/INSU, France, and the ECOS/Conicyt grant C13U01. This research received the support of PHASE, the high angular resolution partnership between ONERA, the Observatoire de Paris, CNRS, and University Denis Diderot Paris 7. This work made use of the SIMBAD and VIZIER astrophysical database from CDS, Strasbourg, France and the bibliographic informations from the NASA Astrophysics Data System. Data processing for this work have been done using the Yorick language which is freely available at http://yorick.sourceforge.net/.
\end{acknowledgements}


\bibliographystyle{aa}   
\bibliography{/Users/alex/Sciences/Articles/bibliographie}

\end{document}